\documentclass[twocolumn,showpacs,preprintnumbers]{revtex4}
\usepackage{amssymb}
\usepackage{amsmath}
\usepackage{graphicx}
\usepackage{dcolumn}
\usepackage{bm}

\begin{document}

\title{Quantum phase transition of light in the dissipative Rabi-Hubbard
lattice: A dressed-master-equation perspective}
\author{ Tian Ye$^{1,2}$}
\author{Chen Wang$^{3,}$}
\email{wangchenyifang@gmail.com}
\author{Qing-Hu Chen$^{1,2,}$}
\email{qhchen@zju.edu.cn}

\address{
$^{1}$ Zhejiang Province Key Laboratory of Quantum Technology and Device, Department of Physics, Zhejiang University, Hangzhou 310027, China \\
$^{2}$  Collaborative Innovation Center of Advanced Microstructures, Nanjing University, Nanjing 210093, China\\
$^{3}$ Department of Physics, Zhejiang Normal University, Jinhua 321004, Zhejiang, China
 }\date{\today }

\begin{abstract}
In this work, we investigate the quantum phase transition of light in the
dissipative Rabi-Hubbard lattice under the framework of the mean-field
theory and quantum dressed master equation. The order parameter of photons
in strong qubit-photon coupling regime is derived analytically both at zero
and low temperatures. Interestingly, we can locate the localization and
delocalization phase transition very well in a wide parameter region. {In
particular for the zero-temperature limit, the critical tunneling strength
approaches zero generally in the deep-strong qubit-photon coupling regime,
regardless of the quantum dissipation. This is contrary to the previous
results with the finite minimal critical tunneling strength based on the
standard Lindblad master equation. Moreover, a significant improvement of
the critical tunneling is also observed at finite temperature, compared with
the counterpart under the Lindblad description. We hope these results may
deepen the understanding of the phase transition of photons in the
Rabi-Hubbard model. }
\end{abstract}

\pacs{42.50.Lc, 42.50.Pq, 05.30.Rt}
\maketitle

\section{\label{Introduction} Introduction \newline
}

The microscopic interaction between light and quantum matter is ubiquitous
in broad fields ranging from quantum optics~\cite%
{mscully1997book,ewaks2018science}, condensed-matter physics~\cite%
{llu2014np,crp2016lehur,tozawa2019rmp}, to quantum chemistry ~\cite%
{rribeiro2018cs,jzhou2019pnas}, and has attracted tremendous attention in
many years. It continues to be a hot topic due to the tremendous progress in
supercondcting qubits ~\cite{tniemczyk2010np,Forn}, trapped ions~\cite%
{trap_ions}, and cold atoms~\cite{cold_atom}. The simplest paradigm is
composed of a two-level system interacting with a single-mode radiation
field, theoretically characterized by the seminal quantum Rabi model~\cite%
{rabi_rabimodel} and its restricted model after the rotating-wave
approximation, known as the Jaynes-Cummings (JC) model~\cite{jcm1963ieee}.
Recently, due to the significant development of circuit quantum
electrodynamics (cQED)~\cite{tniemczyk2010np,fyoshihara2017np}, light-matter
coupling now has reached the ultrastrong and even deep-strong coupling
regimes, which invalidates the rotating-wave approximation and spurs a
plethora of inspiring works~\cite{dressedME,
braak_rabi,chen_rabi,Braak2,USC_RMP,nmueller2020nature,Dicke_intro,mjhwang2015prl,mxliu2017prl}%
.

{\ While considering the interplay between the on-site light-matter
interaction and intersite photon hopping lattice, the {\ Rabi (JC)-Hubbard
model } is the representative cQED lattice model~\cite%
{QPT_JCH,mhartmann2006np,crp2016lehur}. One of the most intriguing effects
for the light-matter interacting lattice is the {\ quantum phase transition
(QPT)} of light, i.e., the localization to delocalization transition of
photons in the ground state. Specifically, multiple Mott-to-superfluid
transitions and a series of Mott lobes are exploited in the JC-Hubbard model~%
\cite%
{QPT_JCH,mhartmann2006np,sclei2008pra,koch2009pra,schmidt2009prl,nietner2012pra,jbyou2014prb,bb2014pra,ahayward2016prb}%
, whereas the Mott lobes are strongly suppressed and a single global
boundary is unraveled to clarify the localized and delocalized phases in the
{\ Rabi-Hubbard model}~\cite{hzheng2011pra,QPT_RH,mschiro2013jpb,yclu2016qip}%
. Meanwhile, the mean-field theory is confirmed to be an efficient and
reliable approach to consistently obtain the phase diagram of photons~\cite%
{QPT_JCH,koch2009pra,QPT_RH}, which reduces the lattice system to the
order-parameter driven single-site model. }

{\ Practically, one quantum system inevitably interacts with the environment~%
\cite{uweiss2012book}. For the dissipative JC-Hubbard model, the effective
non-Hermitian Hamiltonian description leads to the suppression of Mott lobes~%
\cite{rcwu2017pra}. While for the dissipative {\ Rabi-Hubbard model}, Schir%
\'{o} \emph{et al.}~\cite{Fazio_DPT_RH} applied {\ the} Lindblad master
equation (LME) to clarify exotic phases based on the spin-spin correlation
function. In quantum optics, it was reported that as the qubit-photon
interaction becomes strong, the light-matter hybrid system should be treated
as a whole. This implies that the quantum master equation should be
microscopically derived in the eigenspace of the hybrid system, rather than
in the local-component basis (e.g., qubit and resonator). This directly
results in the emergence of generalized master equations (GME) and failure
of the Lindblad description~\cite{dressedME}. Moreover, the finite-time
dynamics based on {\ the} GME shows significant distinction from the
counterpart under {\ the} LME~\cite%
{dressedME,asettineri2018pra,aboite2016aqt}. After long-time evolution, {\
the} GME is usually reduced to the dressed master equation (DME)~\cite%
{asettineri2018pra,lg2015pra,aridolfo2012prl}, where off-diagonal elements
of the density matrix of the hybrid system become negligible due to the full
thermalization. Hence, {\ the} DME can be properly employed to investigate
the steady-state behavior of the hybrid quantum systems from weak to strong
hybridization strengths~\cite%
{asettineri2018pra,aboite2016aqt,dressedME,aridolfo2012prl,alb2017pra,aridolfo2013prl,aboite2016pra}%
. However, to the best of our knowledge, the application of {\ the} DME to
investigate steady-state phase transition of light in the dissipative
Rabi-Hubbard model currently lacks exploration, even under the mean-field
framework. }

\indent In this work, we apply {\ the} DME combined within mean-field theory
to study the steady-state phase diagram of the dissipative Rabi-Hubbard
model. The QPT of light are exhibited both at zero and finite temperatures.
The nontrivial analytical expression of the order parameter is obtained,
which relies on a comprehensive set of system parameters. Moreover, an
improved boundary to characterize {\ the} phase transition of photons is
also achieved, compared with the previous work with {\ the} Lindblad
dissipation~\cite{Fazio_DPT_RH}. The paper is organized as follows: In Sec.
II we briefly introduced the dissipative Rabi-Hubbard model and the DME. In
Sec. III we numerically show the phase diagram of the order parameter of
photons, and analytically locate the phase boundary. Finally, we give a
summary in Sec. IV.

\section{\label{mm}Model and Method}

\subsection{\label{model}The Rabi-Hubbard model}

The Rabi-Hubbard lattice system, which is composed by the on-site quantum
Rabi model and photon tunneling between the nearest-neighboring sites with
the strength $J$, is described as
\begin{equation}
H=\sum_{n}H_{n}^{\text{Rabi}}-J\sum_{{\langle }m,n{\rangle }}(a_{m}^{\dagger
}a_{n}+a_{m}a_{n}^{\dagger }),~  \label{trabi}
\end{equation}%
where $H_{n}^{\text{Rabi}}$ denotes the Hamiltonian of the Rabi model at the
$n$th site, which describes the interaction between a qubit and a
single-mode photon field. The Hamiltonian is specified as
\begin{equation}
~H_{n}^{\text{Rabi}}=\omega _{0}a_{n}^{\dagger }a_{n}+\frac{\varepsilon }{2}%
\sigma _{z}^{n}+g\sigma _{x}^{n}(a_{n}+a_{n}^{\dagger }),  \label{nrabi}
\end{equation}%
where $a_{n}$ and $a_{n}^{\dagger }$ are the annihilation and creation
operator of cavity field at the $n$th site, $\sigma _{x}^{n}$ and $\sigma
_{z}^{n}$ are the Pauli operators of qubit at the nth site, $\omega _{0}$
denotes the frequency of cavity field, $\varepsilon $ is the energy
splitting of qubit, and $g$ is the qubit-photon coupling strength.

Then, we adopt the mean-field theory to simplify the Rabi-Hubbard model to
an effective single-site {\ model}. {\ Specifically, the photon hopping term
in Eq.~(\ref{trabi}) is decoupled as $a_{m}^{\dagger }a_{n}=a_{m}^{\dagger }{%
\langle }a_{n}{\rangle }+{\langle }a_{m}^{\dagger }{\rangle }a_{n}-{\langle }%
a_{m}^{\dagger }{\rangle }{\langle }a_{n}{\rangle }$~\cite{QPT_JCH}. }
Consequently, the Hamiltonian of {\ the } mean-field Rabi model is given by
\begin{eqnarray}
~H_{\text{MF}}(\psi ) &=&\omega _{0}a^{\dagger }a+\frac{\varepsilon }{2}%
\sigma _{z}+g\sigma _{x}(a+a^{\dagger })-zJ\psi (a+a^{\dagger })  \notag
\label{mfrabi} \\
&&+zJ\psi ^{2},
\end{eqnarray}%
where $z$ is the number of {\ } nearest neighboring sites and $\psi ={%
\langle }a{\rangle }$ can be regarded as the order parameter. The subindex $%
n $ is ignored for all sites, because each site shares the same Hamiltonian
in the mean-field framework. The emergence of the nonzero order parameter $%
\psi $ can be used to characterize the QPT of light in the Rabi-Hubbard
lattice, i.e., the localization phase to delocalization phase transition. {\
We note that though the reduced mean-field Rabi model at Eq.~(\ref{mfrabi})
can be efficiently solved at steady state, the mean-field theory indeed has
its own limitations~\cite{QPT_JCH,htc2001pra,htc2003pra}. It includes the
decoupling approximation, i.e., the effective driving strength $zJ\psi $
should be weak. }

Due to the tremendous advance of superconducting circuits engineering, the
ultrastrong qubit-resonator coupling was experimentally detected in cQED~%
\cite{tniemczyk2010np,fyoshihara2017np}, which is theoretically described by
the quantum Rabi model. Meanwhile, several large cQED lattices have also
been designed~\cite{aah2012np,ic2013rmp,mf2017prx,ajk2019nature,ic2020np},
which provide the solid ground to simulate the JC-Hubbard model. Hence, we
believe the Rabi-Hubbard model ~(\ref{trabi}) could be realized by combining
these two components.

\subsection{\label{ME} Quantum dressed master equation}

We take quantum dissipation into consideration in this work. Specifically,
we include local dissipation where the $n$th site mean-field Rabi model is
coupled to two individual bosonic thermal baths. Hence, the total
Hamiltonian under the mean-field theory can be expressed as
\begin{equation}
H_{t}=H_{\text{MF}}(\psi)+H_{B}+V.
\end{equation}%
The bosonic thermal baths are described as $H_{B}=\Sigma _{u=q,c}\Sigma
_{k}\omega _{k}b_{u,k}^{\dagger }b_{u,k}$, where $b_{u,k}$ and $%
b_{u,k}^{\dagger }$ are the annihilation and creation operators of the boson
with the frequency $\omega _{k}$ in the $u$th bath. The interactions between
the Rabi-Hubbard model and bosonic thermal baths are given by $V=V_{q}+V_{c}$%
, {\ where the interaction terms associated with the qubit and the cavity
respectively read}
\begin{eqnarray}
V_{q} &=&\Sigma _{k}\lambda _{q,k}(b_{q,k}+b_{q,k}^{\dagger })\sigma _{x},~
\label{vq} \\
V_{c} &=&\Sigma _{k}\lambda _{c,k}(b_{c,k}+b_{c,k}^{\dagger })(a+a^{\dagger
}),~  \label{vc}
\end{eqnarray}%
with $\lambda _{q,k}(\lambda _{c,k})$ the coupling strength between the
qubit (photon) and the corresponding thermal bath. The system-bath
interaction is characterized as the spectral function $G_{q(c)}(\omega
)=2\pi \sum_{k}|\lambda _{q(c),k}|^{2}\delta (\omega -\omega _{k})$. In this
work, we select the Ohmic case to quantify the thermal bath, i.e., $%
G_{q}(\omega )=\gamma _{q}\omega /\varepsilon \exp (-\omega /\omega _{c})$
and $G_{c}(\omega )=\gamma _{c}\omega /\omega _{0}\exp (-\omega /\omega
_{c}) $, where $\gamma _{q(c)}$ is the dissipation strength and $\omega _{c}$
is the cutoff frequency. $\omega _{c}$ is considered to be large enough, so
the spectral functions are simplified as $G_{q}(\omega )=\gamma _{q}\omega
/\varepsilon $ and $G_{c}(\omega )=\gamma _{c}\omega /\omega _{0}$.

Next, we assume weak coupling between the quantum system and bosonic thermal
baths. {\ Regarding the system-bath interactions ~(\ref{vq}) and~(\ref{vc})
as perturbations, we obtain {\ the} GME under {\ the} Born-Markov
approximation as~\cite{asettineri2018pra,aboite2016aqt}
\begin{eqnarray}
~\frac{\partial {\rho }_{\text{MF}}}{{\partial }t} &=&-i[{H}_{\text{MF}%
}(\psi ),{\rho }_{\text{MF}}]  \notag  \label{qme0} \\
&&+\frac{1}{2}\sum_{\omega ,\omega ^{\prime };u=q,c}\{\kappa _{u}(\omega
^{\prime })[{P}_{u}(\omega ^{\prime }){\rho }_{\text{MF}},{P}_{u}(\omega )]
\notag \\
&&+\mathrm{H.c.}\},
\end{eqnarray}%
where the rate is given by $\kappa _{u}(\omega )=G_{u}(\omega )n_{u}(\omega
) $, with $n_{u}(\omega )=1/[\exp (\omega _{k}/k_{B}T_{u})-1]$ being the
Bose-Einstein distribution function, the projecting operator of the
resonator derived from $[{a}^{\dag }(-\tau )+{a}(-\tau )]=\sum_{\omega }{P}_{%
\text{c}}(\omega )e^{-i\omega \tau }$, is given by ${P}_{c}(\omega
)=\sum_{n,m}{\langle }\phi _{n}|(a^{\dag }+a)|\phi _{m}{\rangle }\delta
(\omega -E_{n,m})|\phi _{n}{\rangle }{\langle }\phi _{m}|$, and the
projecting operator of the qubit, based on the relation ${\sigma }_{x}(-\tau
)=\sum_{\omega }{P}_{q}(\omega )e^{-i\omega \tau }$, is given by ${P}%
_{q}(\omega )=\sum_{n,m}{\langle }\phi _{n}|{\sigma }_{x}|\phi _{m}{\rangle }%
\delta (\omega -E_{n,m})|\phi _{n}{\rangle }{\langle }\phi _{m}|$, where the
energy gap is $E_{n,m}=E_{n}-E_{m}$, with $\hat{H}_{\text{MF}}|\phi _{n}{%
\rangle }=E_{n}|\phi _{n}{\rangle }$. }

After a long-time evolution, the off-diagonal density-matrix elements of the
mean-field Rabi model expressed in the eigenstate representation are
negligible. Then, the populations are decoupled from the off-diagonal
elements, which simplifies the pair of projectors ${P}_{\mu }(\omega )$ and $%
{P}_{\mu }(\omega ^{\prime })$ to ${P}_{\mu }(\omega =E_{n}-E_{m})={\langle }%
\phi _{n}|{A}_{\mu }|\phi _{m}{\rangle }|\phi _{n}{\rangle }{\langle }\phi
_{m}|$ and ${P}_{\mu }(\omega ^{\prime }=E_{m}-E_{n})={\langle }\phi _{m}|{A}%
_{\mu }|\phi _{n}{\rangle }|\phi _{m}{\rangle }{\langle }\phi _{n}|$, with $%
A_{c}=(a^{\dag }+a)$ and $A_{q}=\sigma _{x}$. Then, {\ the} quantum master
equation can be simplified to   the  DME. Specifically, DME is expressed as~
\cite{dressedME,aridolfo2012prl,aridolfo2013prl,aboite2016pra,asettineri2018pra}
\begin{eqnarray}
~\frac{\partial \rho _{\text{MF}}}{\partial t} &=&-i[H_{\text{MF}}(\psi
),\rho _{\text{MF}}]  \notag  \label{mfme} \\
&&+\sum_{j,k>j;u=q,c}\{\Gamma _{u}^{jk}\left( 1+n_{u}\left( \Delta
_{kj}\right) \right) \mathcal{D}[|\phi _{j}\rangle \langle \phi _{k}|,\rho _{%
\text{MF}}]  \notag \\
&&+\Gamma _{u}^{jk}n_{u}\left( \Delta _{kj}\right) \mathcal{D}[|\phi
_{k}\rangle \langle \phi _{j}|,\rho _{\text{MF}}]\},
\end{eqnarray}%
where the dissipator is $\mathcal{D}[O,\rho _{\text{MF}}]=\frac{1}{2}(2O\rho
_{\text{MF}}O^{\dagger }-O^{\dagger }O\rho _{\text{MF}}-\rho _{\text{MF}%
}O^{\dagger }O)$, {$\Delta _{kj}=E_{k}-E_{j}$ is the transition frequency of
two energy levels, and the effective dissipative rates $\Gamma _{q}^{kj}$
and $\Gamma _{c}^{kj}$ thus read
\begin{subequations}
\begin{align}
\Gamma _{q}^{kj}=& \frac{\gamma _{q}\Delta _{kj}}{\varepsilon }|\langle \phi
_{j}|(\sigma _{-}+\sigma _{+})|\phi _{k}\rangle |^{2}, \\
\Gamma _{c}^{kj}=& \frac{\gamma _{c}\Delta _{kj}}{\omega _{0}}|\langle \phi
_{j}|(a+a^{\dagger })|\phi _{k}\rangle |^{2}.
\end{align}
\end{subequations}
\indent With the DME, we can self-consistently solve the steady state of the
Rabi-Hubbard model. To be specific, {we initially set the order parameter to
be an arbitrary reasonable value, and find a temporary steady state $\rho
_{ss}$. Then we calculate the order parameter
\begin{equation}
\psi =\text{Tr}\{\rho_{ss}a\},  \label{order}
\end{equation}%
for the next-step iteration. This procedure can be repeated until the
converged steady state and order parameter are achieved. All physical
quantities can be calculated within the final steady state.}

\subsection{Two-dressed-state approximation}

{\ In the limiting regime of {\ the} deep-strong qubit-photon coupling, low
temperature, and weak excitation of the order parameter, we may confine the
complete Hilbert space to the subspace only spanned by the ground state $%
|\phi _{0}{\rangle }$ and the first-excited state $|\phi _{1}{\rangle }$.
Moreover, we approximately replace the eigenstate $|\phi _{k}{\rangle }$ of
the mean-field Rabi model ~(\ref{mfme}) by the counterpart $|\phi _{k}^{%
\text{Rabi}}{\rangle }$ of the standard Rabi model $\hat{H}^{\text{Rabi}}$
by ignoring $\psi $ ~(\ref{mfrabi}). Thus, The Hamiltonian of mean-field
Rabi model approximates as
\begin{equation*}
H_{\text{MF}}(\psi )\approx \left(
\begin{array}{cc}
\langle \phi _{0}^{\text{Rabi}}| & \langle \phi _{1}^{\text{Rabi}}|%
\end{array}%
\right) \left(
\begin{array}{cc}
0 & \beta _{\psi } \\
\beta _{\psi } & \Delta \\
&
\end{array}%
\right) \left(
\begin{array}{cc}
|\phi _{0}^{\text{Rabi}}\rangle &  \\
|\phi _{1}^{\text{Rabi}}\rangle &
\end{array}%
\right) ,
\end{equation*}%
where the coefficient is $\beta _{\psi }=-zJ\psi \langle \phi _{0}^{\text{%
Rabi}}|(a+a^{\dagger })|\phi _{1}^{\text{Rabi}}\rangle $ and the energy gap
is $\Delta =E_{1}^{\text{Rabi}}-E_{0}^{\text{Rabi}}$. }

The eigenstates under the adiabatic approximation are given by~\cite%
{Irish2005prb}
\begin{subequations}
\begin{align}
~|\psi _{0}^{\text{Rabi}}\rangle \approx & \frac{1}{\sqrt{2}}(|-\rangle
_{x}|\alpha \rangle -|+\rangle _{x}|-\alpha \rangle ),  \label{es1} \\
|\psi _{1}^{\text{Rabi}}\rangle \approx & \frac{1}{\sqrt{2}}(|-\rangle
_{x}|\alpha \rangle +|+\rangle _{x}|-\alpha \rangle ),  \label{wavefun}
\end{align}%
with $\sigma _{x}|\pm \rangle _{x}=\pm |\pm \rangle _{x}$, $|\alpha \rangle $
is the coherent state, and the displaced coefficient $\alpha ={g}/{\omega
_{0}}$. Then, the DME is simplified as
\end{subequations}
\begin{subequations}
\begin{align}
\frac{\partial \rho _{10}}{\partial t}=& -i\beta _{\psi }(2\rho _{00}-1)
\notag \\
& -\frac{\Gamma (\Delta )}{2}[2n_{\text{B}}(\Delta )+1]\rho _{10}-i\Delta
\rho _{10},  \label{dme_ana1} \\
\frac{\partial \rho _{00}}{\partial t}=& i\beta _{\psi }(\rho _{01}-\rho
_{10})  \notag \\
& +\frac{\Gamma (\Delta )}{2}\{1-[2n_{\text{B}}(\Delta )+1](2\rho
_{00}-1)\},~  \label{dme_analytic}
\end{align}%
where the reduced density-matrix element is $\rho _{ij}={\langle }\phi _{i}^{%
\text{Rabi}}|\hat{\rho}_{s}(t)|\phi _{j}^{\text{Rabi}}{\rangle }~(i,j=0,1)$,
the dissipation rate becomes $\Gamma (\Delta )={\gamma _{c}\Delta |\langle
\phi _{0}^{\text{Rabi}}|(a+a^{\dagger })|\phi _{1}^{\text{Rabi}}\rangle |^{2}%
}/{\omega _{0}}+{\gamma _{q}\Delta |\langle \phi _{0}^{\text{Rabi}}|(\sigma
_{-}+\sigma _{+})|\phi _{1}^{\text{Rabi}}\rangle |^{2}}/{\varepsilon }$, and
the Bose-Einstein distribution function is $n_{\text{B}}(\Delta )=1/[\exp
(\Delta /k_{B}T)-1]$, with $T_{q}=T_{c}=T$. Here, the order parameter can be
reexpressed as $\psi =g(\rho _{01}^{ss}+\rho _{10}^{ss})/\omega _{0}$ and $%
\beta _{\psi }=-2g^{2}zJ(\rho _{01}^{ss}+\rho _{10}^{ss})/\omega _{0}^{2}$,
with $\rho _{10}^{ss}$ and $\rho _{01}^{ss}$ being the density elements at
steady state. Moreover, based on the eigenstates ~(\ref{es1}) the energy gap
and the transition rate can be obtained as~\cite{mschiro2013jpb}
\end{subequations}
\begin{eqnarray}~\label{delta1}
\Delta \approx\varepsilon \exp (-2g^{2}/\omega _{0}^{2})
\end{eqnarray}
and
\begin{eqnarray}~\label{gammad1}
\Gamma(\Delta){\approx }\Delta \left( \frac{4g^{2}\gamma _{c}}{\omega_{0}^{3}}+\frac{\gamma _{q}}{\varepsilon }\right),
\end{eqnarray}
respectively.

\section{Results and discussions}

\subsection{Numerical analysis of QPT}

\begin{figure}[tbp]
\centering	\includegraphics[width=0.5\textwidth]{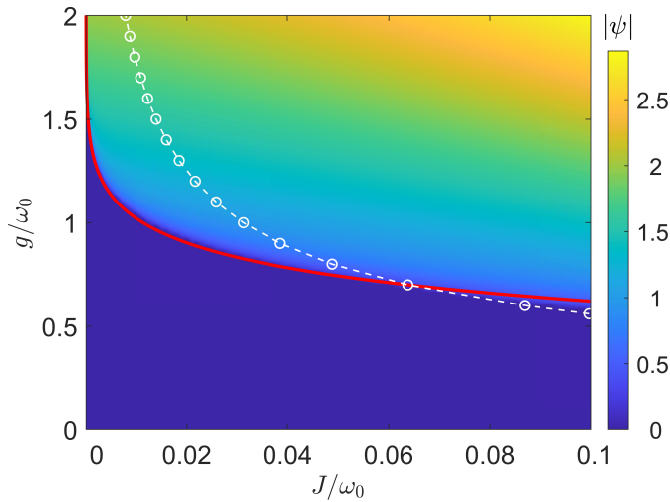}
\caption{(color online). The steady state phase diagram of the order
parameter $|\protect\psi |$ at zero temperature. The red solid line shows
the phase boundary described by the analytical solution at Eq.~(\protect\ref%
{zJc}), and the white dashed curve with circles shows the boundary based on
the LME~\protect\cite{Fazio_DPT_RH}. The other system parameters are given
by $\protect\varepsilon =\protect\omega _{0}$, z=3, $\protect\gamma _{q}=%
\protect\gamma _{c}=10^{-4}\protect\omega _{0}$. {\ The selected system
parameters are consistent with previous works, e.g., $\protect\omega_0/2%
\protect\pi=\protect\varepsilon/2\protect\pi=6$~GHz, $\protect\gamma_q/%
\protect\omega_0=\protect\gamma_c/\protect\omega_0{=}0.2{\times}10^{-4}$ in
Ref.~\protect\cite{dressedME}; $\protect\omega_0/2\protect\pi=4$~GHz, $%
\protect\varepsilon/2\protect\pi=8$~GHz, $\protect\gamma_q/\protect\omega_0=%
\protect\gamma_c/\protect\omega_0=1.8{\times}10^{-4}$ in Ref.~\protect\cite%
{lg2015pra}. } }
\label{fig_RL_PD}
\end{figure}

We first apply {\ the} DME combined with mean-field theory to numerically
investigate the phase diagram of photons of the Rabi-Hubbard model at zero
temperature. {\ By calculating the order parameter $|\psi |$} with Eq.~(\ref%
{order}), the sharp phase transition of photons from the localization to the
delocalization phase is clearly exhibited in Fig.\ref{fig_RL_PD}. {The
localization phase with vanished order parameter (i.e., $|\psi |=0$) is
denoted by the dark blue region, whereas the delocalization phase,
characterized as significant excitation of the order parameter, locates at
light green and yellow regions. The present phase transition corresponds to
the $Z_{2}$ symmetry breaking}. As the inter-site photon tunneling strength $%
J$ increases, the critical on-site qubit-photon coupling strength $g_{c}$
separating the delocalization and localization phases decreases gradually. {%
\ In particular, as $g_c$ approaches $\infty$, it is found that the
corresponding $J_c{\rightarrow}0$. We also study the phase boundary at
finite temperature, e.g., $T=0.05\omega_0$ in Fig.~\ref{fig_RL_PD_T005}. The
delocalization phase is partially suppressed, particularly in the weak
photon tunneling regime, which leads to finite $J_c$ for $g_c{\rightarrow}%
\infty$. Hence, {\ the} thermal noise favors the localization phase of
photons. }

These numerical results are qualitatively consistent with the previous ones
observed in the QPT of light at both the steady state~\cite{QPT_RH} and the
ground state~\cite{Fazio_DPT_RH}, where the competition between the
inter-site photon tunneling and the on-site qubit-photon coupling are also
considered. In the next subsection, we analytically explain the global
boundary through the order parameter of photons.

\begin{figure}[tbp]
\centering	\includegraphics[width=0.5\textwidth]{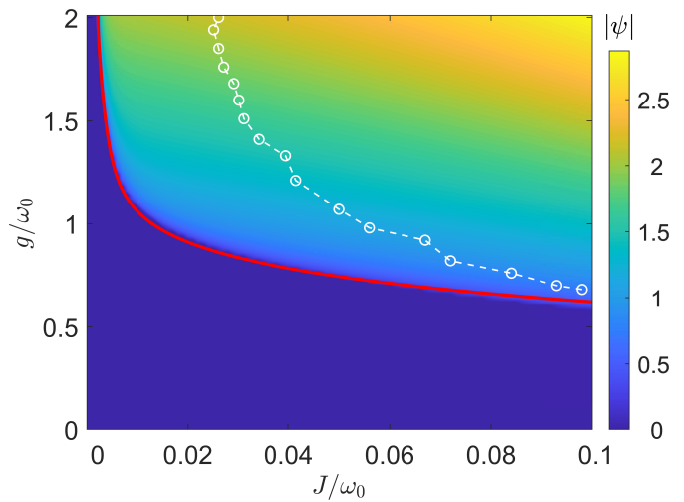}
\caption{(color online). The steady state phase diagram of the order
parameter $|\protect\psi |$ at finite temperature, $T_q =T_c =0.05\protect%
\omega_{0}$. The red solid line shows the phase boundary described by the
analytical solution at Eq.~(\protect\ref{zjcT}), and the white dashed curve
with circles shows the boundary based on the LME numerically. The other
system parameters are the same as those in Fig.~\protect\ref{fig_RL_PD}.}
\label{fig_RL_PD_T005}
\end{figure}

\subsection{Analytical solution of quantum phase transition}

{\ Since the phase boundary located by numerical calculation in Fig.~\ref%
{fig_RL_PD} is just in the deep-strong coupling regime where the analytical
adiabatic approximation can be applicable \cite{Irish2005prb}, we may also
derive an analytical solution of the order parameter $|\psi |$ at steady
state in the critical regime. Specifically, at steady state ($d\rho
_{ij}^{ss}/dt=0$) we obtain the {\ following} relations from Eqs.~(\ref%
{dme_ana1})-(\ref{dme_analytic})
\begin{subequations}
\begin{align}
\text{Im}\{\rho^{ss}_{10}\}=&\Gamma(\Delta)[n_\text{B}(\Delta)+1/2]\text{Re}%
\{\rho^{ss}_{10}\}/\Delta,~  \label{im10} \\
1-2\rho^{ss}_{00}=&({\Gamma^2(\Delta)[n_\text{B}(\Delta)+1/2]^2}/{\Delta}%
+\Delta)\text{Re}\{\rho^{ss}_{10}\}/\beta_\psi~ .  \label{re10}
\end{align}
In the zero temperature limit,  combining  Eq. (\ref{dme_analytic}) with Eqs. ~(\ref{im10}%
)  and~(\ref{re10}), the order parameter is obtained as
\end{subequations}
\begin{eqnarray}
~ |\psi |\approx \frac{\omega _{0}}{2\sqrt{2}gzJ}\sqrt{\frac{4g^{2}zJ\Delta
}{\omega _{0}^{2}}-\Delta ^{2}-\frac{\Gamma ^{2}(\Delta )}{4}}.
\label{psiT0}
\end{eqnarray}
Moreover, from Eq.~(\ref{psiT0}) it is explicitly shown that the emergence
of the nonzero order parameter is bounded by the inequality
\begin{equation}
~\frac{4g^{2}zJ\Delta }{\omega _{0}^{2}}{\geq }\Delta ^{2}+\frac{\Gamma
^{2}(\Delta )}{4}.  \label{ine1T0}
\end{equation}%
Such inequality has pronounced consequence on the phase-transition boundary
in Fig.~\ref{fig_RL_PD}. Then, we analyze the boundary of QPT of photons
from the analytical perspective. }

In one previous study of the dissipative Rabi-Hubbard lattice~\cite%
{Fazio_DPT_RH}, Schir\'{o} et al. applied the LME to approximately quantify
the phase boundary with the critical tunneling strength at zero temperature $%
J_{\text{crit}}{\approx }[{\gamma _{c}^{2}g^{2}}/{\omega _{0}^{3}}+{\omega
_{0}^{3}}/{16g^{2}}]/d$, with $d$ being the dimensional number, which is
also reproduced by the white dashed curve with circles in Fig.~\ref%
{fig_RL_PD}. In the photon-tunneling regime, $J_{\text{crit}}>0.06\omega
_{0} $, their result agrees with the numerical counterpart. However, for the
strong qubit-photon coupling, one can note that $J_{\text{crit}}$ has a
finite minimal value in Ref.~\cite{Fazio_DPT_RH}, i.e., $\gamma _{c}/2d$.
This result is inconsistent with the present numerical result in Fig.~\ref%
{fig_RL_PD} and those observed in the Rabi-Hubbard model in absence of the
quantum dissipation~\cite{QPT_RH}, in which $J_{\text{crit}}{\propto }\Delta
$ vanishes with the increase of $g$.

Here, considering the inequality ~(\ref{ine1T0}) with $\Delta $ and $\Gamma
(\Delta )$ specified by Eqs.~(\ref{delta1}) and (\ref{gammad1}), we obtain
the critical tunneling strength as
\begin{equation}
~zJ_{c}=\frac{\omega _{0}^{2}\Delta }{4g^{2}}\left[ 1+\Big{(}\frac{%
2g^{2}\gamma _{c}}{\omega _{0}^{3}}+\frac{\gamma _{q}}{2\varepsilon }\Big{)}%
^{2}\right] .  \label{zJc}
\end{equation}%
It is found that in the absence of quantum dissipation, i.e., $\gamma
_{q(c)}=0$, $J_{c}$ is naturally reduced to ${\omega _{0}^{2}\Delta }/{%
(4zg^{2})}$, which is quite compatible with the result in Ref.~\cite{QPT_RH}%
. While by tuning on the dissipation, the critical tunneling strength is
shifted up by the cooperative contribution of the dissipation strengths,
i.e., $\gamma _{q}$ and $\gamma _{c}$. However, as the qubit-photon coupling
becomes deep strong, $J_{c}$ again approaches zero, as indicated by the red
curve in Fig.~\ref{fig_RL_PD}. We immediately note that this result at Eq.~(%
\ref{zJc}) is different from that based on the LME in Ref.~\cite%
{Fazio_DPT_RH} in the strong qubit-photon interaction regime.

Physically, {\ the } DME captures the microscopic transitions from the
eigenstate $|\phi _{1}^{\text{Rabi}}{\rangle }$ to $|\phi _{0}^{\text{Rabi}}{%
\rangle }$, which is characterized as the rate $\Gamma (\Delta )$ ~(\ref%
{gammad1}). Hence, the dynamical transition in {\ the } DME generally relies
on the energy gap, e.g., by selecting Ohmic (in this work) or super-Ohmic
types of thermal baths, which leads to the result that $J_{c}{\rightarrow }0$
as $g_{c}{\rightarrow }\infty $. In contrast, LME phenomenologically treats $%
\Gamma (\Delta )$ independent of the energy gap, which generally
overestimates the dissipative processes at strong qubit-photon coupling.
Therefore, we believe that {\ the } DME may generally improve the boundary
from the microscopic view.

{\ We also analyze the phase boundary at finite temperatures. From Eqs.~(\ref%
{dme_analytic}), (\ref{im10}), and (\ref{re10}), the order parameter $|\psi
| $ can be obtained as 
\begin{eqnarray}
|\psi | &\approx &\sqrt{\frac{4g^{2}zJ\Delta }{(2n_{\text{B}}(\Delta
)+1)\omega _{0}^{2}}-\Delta ^{2}-\frac{\left[ (2n_{\text{B}}(\Delta
)+1)\Gamma (\Delta )\right] ^{2}}{4}}  \notag  \label{psi_Tn0} \\
&&{\times }\frac{\omega _{0}}{2\sqrt{2}gzJ}.
\end{eqnarray}%
Meanwhile, the nonzero order parameter is bounded by the inequality
\begin{equation}
~\frac{4g^{2}zJ\Delta }{[2n_{\text{B}}(\Delta )+1]\omega _{0}^{2}}{\geq }%
\Delta ^{2}+\frac{[2n_{\text{B}}(\Delta )+1]^{2}\Gamma ^{2}(\Delta )}{4}.
\label{ine1}
\end{equation}%
Hence, we may predict the critical tunneling strength at finite temperature
as
\begin{eqnarray}
~zJ_{c} &=&{\frac{\omega _{0}^{2}\Delta }{4g^{2}}}[2n_{\text{B}}(\Delta
)+1]\{1+[2n_{\text{B}}(\Delta )+1]^{2}  \notag  \label{zjcT} \\
&&\times ({2g^{2}\gamma _{c}}/{\omega _{0}^{3}}+{\gamma _{q}}/{2\varepsilon }%
)^{2}\},
\end{eqnarray}%
which is naturally reduced to Eq.~(\ref{zJc}) as $T$ approaches aero. From
Fig.~\ref{fig_RL_PD_T005} it is found that $J_{c}$ (the red solid line)
based on {\ the } DME agrees well with the numerical result in a wide
qubit-photon coupling regime. Hence, the analytical expression of critical
tunneling strength at Eq.~(\ref{zjcT}) may be helpful to deepen the
understanding of the finite temperature phase transitions of photons.
However, we should admit that this analytical result deviates from the
numerical one at extremely strong qubit-photon coupling limit ($\Delta {%
\approx }0$), where $J_{c}{\propto }g^{2}\exp (4g^{2}/\omega _{0}^{2})$ is
significantly enhanced as $n_{\text{B}}{\approx }k_{B}T/\Delta $. Moreover,
the two-dressed-state approximation may break down at high-temperature
regime, and more dressed states should be considered like in the numerical
analysis. }

\section{\label{conclusion} Conclusion}

In this paper, we study the QPT of light in the Rabi-Hubbard lattice with
local dissipation under the framework of the mean-field theory and the
quantum dressed master equation. The steady-state phase diagram of photons
are numerically calculated, which clearly classifies the localization and
delocalization phases. We then analytically obtain an approximate expression
of the order parameter in the low temperature and deep-strong qubit-photon
coupling regime, where the mean-field Rabi model is reduced to the effective
nearly degenerate two-dressed-state system. We further analyze the boundary
between two difference phases at zero temperature, which is characterized as
the critical tunneling strength. {\ The expression of the critical tunneling
strength in the absence of the quantum dissipation can be naturally reduced
to the previous ones, see, e.g., Ref.~\cite{Fazio_DPT_RH}. While by tuning
on the quantum dissipation, which is characterized by representative thermal
baths, e.g., Ohmic or super-Ohmic type, it is found that the critical
tunneling strength approaches zero as the qubit-photon coupling strength
becomes deep-strong. This result is generally distinct from the previous
work based on {\ the LME}, which in contrast has finite minimal critical
tunneling strength. We also predict the critical tunneling strength at
finite temperatures, which may be helpful to study the phase transition of
light at thermal equilibrium. In the future, it should be interesting to
explore the steady-state phase diagram of photons in the Rabi-Hubbard model
beyond the simple mean-field framework, e.g., based on the cluster
mean-field theory~\cite{jsjin2016prx} and linked-cluster expansion approach~%
\cite{ab2018prb}. }

\section{Acknowledgements}

T. Y and Q.-H.C. are supported by the National Science Foundation of China
under Grant No. 11834005, the National Key Research and Development Program
of China under Grant No. 2017YFA0303002. C.W. acknowledges the National
Natural Science Foundation of China under Grant No. 11704093 and the Opening
Project of Shanghai Key Laboratory of Special Artificial Microstructure
Materials and Technology.

\end{document}